\documentstyle[11pt,aaspp4]{article}

\newcommand{\gtsim}{\ {\raise-0.5ex\hbox{$\buildrel>\over\sim$}}\ }
\newcommand{\ltsim}{\ {\raise-0.5ex\hbox{$\buildrel<\over\sim$}}\ }

\begin{document}

\vspace{12pt}

\title{The disruption of nearby galaxies by the Milky Way}

\vskip 6pt
\author{M. E. Putman$^1$, B. K. Gibson$^1$, L. Staveley-Smith$^2$,
G. Banks$^3$, D. G. Barnes$^4$, R. Bhatal$^5$, 
M. J. Disney$^3$, R. D. Ekers$^2$, K. C. Freeman$^1$, R. F. 
Haynes$^2$,
P. Henning$^6$, 
H. Jerjen$^1$, V. Kilborn$^4$, B. Koribalski$^2$, 
P. Knezek$^7$, D. F. Malin$^8$, J. R. Mould$^1$, T. 
Oosterloo$^9$,
R. M. Price$^2$, S. D. Ryder$^{10}$, E. M. Sadler$^{11}$, 
I. Stewart$^2$, F. Stootman$^5$, R. A. Vaile$^{5\dag}$, 
R. L. Webster$^4$, A. E. Wright$^2$}

\vskip 6pt
\noindent $^1$Mount Stromlo \& Siding Spring Observatories, Australian
National University, Weston Creek P.O., Weston, ACT 2611, Australia

\noindent $^2$Australia Telescope National Facility, CSIRO, P.O. Box 76,
Epping, NSW 2121, Australia

\noindent $^3$University of Wales, Cardiff, Department of Physics \& Astronomy,
P.O. Box 913, Cardiff CF2 3YB, Wales, United Kingdom

\noindent $^4$University of Melbourne, School of Physics, Parkville,
Victoria 3052, Australia

\noindent $^5$University of Western Sydney Macarthur, Department of Physics,
P.O. Box 555, Campbelltown, NSW 2560, Australia

\noindent $^6$University of New Mexico, Department of Physics \& Astronomy, 800
Yale Blvd. NE, Albuquerque, NM 87131, USA

\noindent $^7$The Johns Hopkins University, Department of Physics \& Astronomy,
34$^{\rm th}$ \& North Charles Streets, Baltimore, MD 21218, USA

\noindent $^8$Anglo-Australian Observatory, P.O. Box 296, Epping, NSW 2121,
Australia

\noindent $^9$Istituto di Fisica Cosmica, via Bassini 15, I-20133, Milano, 
Italy 

\noindent $^{10}$Joint Astronomy Center, 660 North Aohoku Place, Hilo, 
HI 96720, USA

\noindent $^{11}$University of Sydney, Astrophysics Department, School of
Physics, A28, Sydney, NSW 2006, Australia

\noindent $^{\dag}$deceased

\newpage
\renewcommand{\baselinestretch}{2}

\bf
  Interactions between galaxies are common and are an important
  factor in determining their physical properties such as
  position along the Hubble sequence and star-formation
  rate$^{1}$.  There are many possible galaxy interaction
  mechanisms, including merging, ram-pressure stripping, gas
  compression, gravitational interaction and cluster tides$^{2}$.
  The relative importance of these mechanisms is often not clear, as
  their strength depends on poorly known parameters such as the
  density, extent and nature of the massive dark halos that surround
  galaxies$^{3}$.  A nearby example of a galaxy interaction where
  the mechanism is controversial is that between our own Galaxy and
  two of its neighbours -- the Large and Small Magellanic Clouds. Here
  we present the first results of a new \ion{H}{1} survey which
  provides a spectacular view of this interaction.  In addition to the
  previously known Magellanic Stream$^{4}$, which trails 100$^{\circ}$
  behind the Clouds, the new data reveal a counter-stream which lies
  in the opposite direction and leads the motion of the Clouds. This
  result supports the gravitational model in which leading
  and trailing streams are tidally torn from the body of the
  Magellanic Clouds. \rm

\vskip 12pt

The Magellanic Stream, discovered 25 yrs ago$^{4}$, is a narrow ($\sim
10^\circ$ wide) tail of neutral hydrogen, which trails the Magellanic
Clouds along a circle around the Milky Way.  The past two decades have
seen the number of viable mechanisms for its formation reduced to two
- \it ram-pressure stripping \rm of material from the Magellanic
System during its passage through the Galactic halo or extended
ionised disk$^{5-8}$, and \it tidal interaction \rm of the Magellanic
Clouds with the Milky
Way$^{9-14}$.  Although tidal models have been successful in
explaining many characteristics of the Stream, they have suffered from
two major drawbacks: (a) the absence of stars in the Stream$^{15}$
(stars are also affected by tidal forces), and (b) the lack of
evidence for a leading stream, or arm, which is a natural
consequence of tidal interaction$^{9-12}$.  Discrete neutral hydrogen
clouds on the leading side have been known for some time, but have
generally been regarded as a separate group of high velocity
clouds$^{5,6}$.

The \ion{H}{1} Parkes All-Sky Survey (HIPASS)$^{16}$ is a new survey
for \ion{H}{1} in the southern sky ($\delta\le 0^\circ$) over the
velocity range -1,200 to 12,700 km s$^{-1}$. It therefore spans the
Milky Way, the Magellanic Clouds and the more distant Universe.  The
64-m Parkes telescope, with a focal-plane array of 13 beams, set in a
hexagonal grid, is being used to survey the sky in $8^\circ$
zones of declination, with approximate Nyquist sampling.  The
spectrometer has 1024 channels for each polarisation and beam, with a
velocity spacing of 13.2 km s$^{-1}$ and a spectral resolution, after
Hanning smoothing, of 26.4 km s$^{-1}$. Whilst not ideal for resolving
fine spectral structure, accurate column densities and velocity fields
are obtained. The HIPASS survey will scan the entire southern sky five
times for full sensitivity. The data presented here come from the
first scan only and cover the sky south of declination $-62^{\circ}$.
The rms brightness temperature sensitivity is approximately 20 mK,
corresponding to a column density sensitivity of $10^{18}$ atoms
cm$^{-2}$ in each channel.  The data were reduced using a modified
version of the standard HIPASS reduction software$^{17}$ (which was
designed for imaging discrete \ion{H}{1} sources).  The bandpass
correction was calculated for each beam and velocity by breaking each
$8^{\circ}$ scan into five sections, finding the median emission in each
section and using the minimum of the five values.  This greatly increases
our sensitivity to large-scale structure without significant loss of
flux density, except near the Galactic Plane.

Figure 1 is a detailed image of the Magellanic System.  This image
shows the \ion{H}{1} peak intensity distribution in a 2400 square
degree mosaic, centred on the South Celestial Pole.  Components such
as the Large Magellanic Cloud (LMC), Small Magellanic Cloud (SMC),
Magellanic Bridge, Galactic Plane and beginning of the Magellanic
Stream are identified.  The feature we wish to focus on here however, lies
``between'' the Clouds and the Galactic Plane and is labelled
``Leading Arm''.  The Leading Arm is shown in more detail in the
channel maps of Figures 2 and 3.
Of particular interest is the connection of the Leading Arm to
the Magellanic Clouds shown in Figure 2 and, despite
its relative thinness ($\sim 1/4$ the width of the trailing Stream), 
the continuity of the Arm to at least Galactic
latitude $b\sim -8^\circ$.  The link
between the Magellanic System and the strong emission features at
$(\ell,b)=(297^\circ,-24^\circ)$ and $(\ell,b)=(302^\circ,-16^\circ)$, is not
visible in the earlier data of Mathewson \& Ford$^{18}$ (their figure
2) or Morras$^{19}$, due to their sparse spatial sampling.  Indeed, it is 
the lack of continuity in the earlier data that was
offered as evidence for the absence of a leading arm$^{6,8}$.  

The velocity distribution of the leading \ion{H}{1} also suggests
a continuous flow of material which originates from the Magellanic
System.  Figure 2 steps through the velocity channels from $v_{\rm lsr}=171$ 
to 356 km s$^{-1}$.  The \ion{H}{1} emission is first seen at the position of the SMC 
 and then continues into the Bridge and the LMC.  There
is subsequently a flow in velocity towards the Galactic
Plane through the newly discovered Leading Arm, with the emission
disappearing at $\sim$ 356 km s$^{-1}$.  In this direction, $(\ell,b)=(300^\circ,0^\circ)$,
the maximum Galactic disk velocity allowed under the assumption of circular
rotation is $\sim 120$ km s$^{-1}$ $^{20}$, in agreement with the observed upper
limit$^{21}$.  The observed velocity of the Leading Arm ($\sim310$ km s$^{-1}$; 
see Figures 2 and 3), coupled with its velocity-space
continuity with the Magellanic System, clearly eliminates the possibility of
the material being associated with the Milky Way disk. 

This continuous Leading Arm, extending $\gtsim 25^\circ$ from an area
near the LMC towards the Galactic Plane, is a natural prediction of tidal
models$^{9-14}$.  The most sophisticated of such models to date$^{10}$
predicts an $\sim 1.5$ Gyr-old leading arm, with a mass $\sim 1/3$ 
that of the Stream, emanating from the SMC/Bridge region
and spanning the region $\ell\approx 280^\circ$ to $310^\circ$ and
$b\approx -30^\circ$ to $+50^\circ$.  When considering the region $b < 0$ (i.e. the 
region shown in Figures 1 - 3), the mass of the predicted arm is $\sim 1/8$ that 
of the Stream, with a relatively flat velocity distribution and a deviation
from the Great Circle defined by the trailing Stream of $\sim30^\circ$.
Some care must be taken when directly comparing these model results
with new observational data, as three-dimensional spatial information
does not exist.  Under the assumption of constant and equal distances, the mass ratio 
of the observed Leading Arm to the Stream is $\sim 1/25$.  However, there is reason
to believe that the tip of the Stream is closer than its head$^{11,12}$, implying
that the inferred observed mass ratio obtained here is actually a lower limit.
The velocity and projected orientation of the observed Leading Arm lie within $\sim50$
km s$^{-1}$ and $\sim30^\circ$, respectively, of that predicted by
Gardiner and Noguchi$^{10}$.   This overall agreement is remarkable when you
consider the simplifying assumptions employed and the limited parameter
space covered (in particular, the shape of the LMC potential and the mass of the 
Galactic Halo).
Our recognition of the missing leading arm feature emphasizes the need to advance 
current tidal models and address the remaining discrepancies.

 
The identification of a continuous Leading Arm to the Stream, of \it
at least \rm $25^\circ$ in length, argues against the ram pressure
model, but with two caveats.  First, in any tidal model, stars are
expected to be torn from the Magellanic System together with the gas,
yet so far searches for stars associated with the Stream have been
negative$^{15,22}$. This may be due to previous searches not being
sensitive to an enhancement of stellar populations older than the
$\sim 1.5$ Gyr Stream age favoured by current tidal models$^{10,23}$.
More probably it is due to the initial gas distribution being
significantly more extended than the stellar component as confirmed in
interferometer studies of spiral galaxies$^{24}$, and as seen in
other interacting systems$^{25}$. A second caveat is that some ram
pressure models form leading features$^{7,8}$; but only at the expense
of significantly worsening the predicted extent and velocity profile
of the Magellanic Stream.

In Figure 4, we present an expanded view of the LMC/SMC
system.  This view is remarkable in showing the LMC to have a much
more regular spiral structure in neutral hydrogen than it does
optically$^{26}$. 
There is also evidence for a thin
tidal tail emanating from the LMC at $(\ell,b)=(284^\circ,-33^\circ)$
and extending into the Bridge region towards the SMC.  Tentative
evidence for such a tail has been presented before$^{27}$, but Figure
4 provides unequivocal evidence for such a tail and thus for a
tidal interaction between the LMC and SMC.  This is in addition to the
evidence presented in Figures 1 - 3 for a tidal interaction between
the Galaxy and the Magellanic System as a whole.

In the future, it will be interesting to determine the full extent of
the Leading Arm. Does there exist a complete ring of Magellanic Cloud
material? A recent metallicity determination$^{28}$ for a high-velocity cloud
lying above the Galactic Plane at $(\ell,b)=(287^{\circ},23^{\circ})$ suggests
a Magellanic Cloud origin, possibly implying a greater extent for the
Leading Arm.  Exploring the velocity structure of the Leading Arm in
more detail will also aid our understanding of the Milky Way's interaction
with the Magellanic Clouds.  These types of
observations provide crucial constraints on simulations aimed at
reproducing the formation and evolution of the Magellanic System. This
remains one of the only extragalactic systems for which an approximate
three-dimensional orbit has been measured and it is an invaluable
probe of the mass and extent of the Galactic halo.  A full
understanding of the role of accretion and tidal interaction in this
local ``Rosetta Stone'' is essential if we are to fully appreciate
their effects in the Universe at large.

\noindent
\bf Acknowledgements. \rm
We acknowledge Warwick Wilson, Mal Sinclair and their teams at CSIRO for their 
engineering
excellence and thank the aips++ team for their software support.  
We also acknowledge: Erwin de Blok, Anne Green, Sebastian Juraszek, Mike 
Kesteven, Rene\'{e} Kraan-Korteweg, Anja Schr\"{o}der, and Lance Gardiner.  

\noindent
\bf Correspondence \rm should be addressed to M.E.P. 
(e-mail putman@mso.anu.edu.au).

\newpage
\renewcommand{\baselinestretch}{2}
\centerline{\bf Figure Captions}

\vskip 12pt
\noindent{\bf Figure 1. \rm The 2400 square degree mosaic of the South
Celestial Pole with the main features labelled, including the area 
containing the newly identified Leading Arm. The Stream extends
100$^\circ$ beyond its labelled starting point.
This image is a peak neutral hydrogen intensity map using the first scans
of HIPASS data (see text) in the velocity range $v_{\rm lsr}\approx 82$ to 
$400$ km s$^{-1}$.  Declination -90$^\circ$ is
at the centre with the edge of the image extending to declination -62$^\circ$.
Right Ascension 0$^{\rm h}$ is at the top of the figure and increases 
anticlockwise.
Note the lack of \ion{H}{1} emission in the
region between the SMC and Leading Arm.  Discontinuity of emission
at the Galactic Plane is an artefact of the bandpass removal. }

\vskip 4pt

\noindent{\bf Figure 2. \rm Channel maps of the Magellanic System
as labelled in Figure 1.   $v_{\rm lsr}$ is labelled in 
km s$^{-1}$ in the left upper corner of each channel.  Note the
continuous flow of the \ion{H}{1} emission starting from the SMC and moving
into the Bridge, the LMC and finally, the newly discovered Leading Arm.
The background striping is a result of the scanning method used for the HIPASS 
survey.
The high positive velocity of the arm eliminates the possibility of stray 
radiation effects$^{29}$ and an association with material from the Milky Way disk$^{20,21}$. 
For comparison, the Magellanic Stream extends from 100 km s$^{-1}$ near the
Clouds, to $-200$ km s$^{-1}$ at its tip$^{4}$.  
The orientation of this figure was chosen to match Figure 1.}

\vskip 4pt

\noindent{\bf Figure 3. \rm A detailed view of the Leading Arm at the channel centred upon
$v_{\rm lsr} = 323$ km s$^{-1}$.
Contours are from 10 - 90\% of the brightness temperature maximum 
(T$_{\rm B}=0.88$ K).
Contrast the continuous nature of the arm, with the previously assumed
``discrete'' nature of the \ion{H}{1} in the Mathewson et~al.$^{17}$ figure 2. 
The total \ion{H}{1} mass of the Arm is $\sim 1\times 10^{7}$ M$_\odot$,
and it deviates from the Great Circle defined by the trailing Stream by $\sim
60^\circ$.  Both the deviation angle and the \ion{H}{1} mass are consistent
with the model predictions of Gardiner \& Noguchi$^{10}$.
The orientation of this figure was chosen to match Figure 1.
}

\vskip 4pt

\noindent{\bf Figure 4. \rm Expanded view of the LMC+SMC+Bridge regions from 
Figure 1.  The LMC is at the bottom left, with the Bridge extending 
to the position of the SMC in the upper right of the figure.  
Note the emission filament emanating directly from the LMC (at
$(\ell,b)=(284^\circ,-33^\circ)$) into the Bridge.  This suggests that the 
Bridge
may contain more mass from the LMC than was previously thought$^{11}$.
The orientation of this figure was chosen to match Figure 1.}

\end{document}